\documentclass{elsart}
\usepackage{harvard}
\usepackage{graphicx}
\usepackage{amssymb}

\def\url#1{{\ttfamily\def\/{/\discretionary{}{}{}}#1}}

\begin{document}
\begin{frontmatter}
\title{Testing minimum energy with powerful radio sources in clusters of 
galaxies}
\author[Leahy]{J. P. Leahy\thanksref{jpl}},
\author[Gizani]{Nectaria A. B. Gizani\thanksref{ng}}
\thanks[jpl]{On sabbatical leave from the University of Manchester.
E-mail: jpl@jb.man.ac.uk}
\thanks[ng]{Present address: University of
Athens, Dept.\ of Physics, Section of Astrophysics, Astronomy and
Mechanics, GR-15784 Zografos, Athens, Greece. 
E-mail: ngizani@cc.uoa.gr}
\address[Leahy]{Space Telescope Science Institute,3700 San Martin Dr., 
Baltimore, MD 21218}
\address[Gizani]{University of Ioannina, Dept.\ of Physics, Section of
Astrogeophysics, 45110 Ioannina, Greece.} 

\begin{abstract}
We analyze {\em ROSAT\/} data for cluster gas surrounding powerful radio
galaxies, which is well fitted by a
``$\beta$-model'' gas distribution, after allowing for a compact central
source.  The cluster thermal pressure at the
distance of the radio lobes is typically an order of magnitude larger than
the lobe minimum pressure. Since radio lobes are sharply-bounded, the 
missing pressure is not simply entrained intra-cluster gas.
Thus the minimum energy in the lobes is a severe underestimate of the actual 
energy content.
We argue that the extra energy is mostly in the form of particles,
so that the magnetic field is below equipartition and thus not a major
factor in the lobe dynamics. The large departure from minimum energy
has far-reaching implications for the nature of AGN central
engines and the supply of mechanical energy to the cluster gas.
\end{abstract}

\end{frontmatter}

\section{Introduction}
\label{intro}

Despite the well-known drawbacks, minimum energies are
the foundation for estimates of the energetics of AGN jets and
the double-lobed radio sources (DRAGNs) they create. The lobes, lying
$\sim 100$ kpc from the centre of large elliptical galaxies, must
be surrounded by gas at $10^7$--$10^8$ K (for simplicity, 
the ``intra-cluster medium''). The ICM confines the lobes either
directly by thermal pressure or, as in conventional models for FR\,IIs,
by ram pressure.  Hence the ambient pressure $P_{\rm th} = 2.3 n_H kT$
gives a lower limit for the lobe energy density, to be
compared with the minimum energy estimate.

In several twin-jet FR\,Is, $u_{min}/3 \equiv P_{\rm min} \ll P_{\rm th}$, 
but these jets are
believed to be strongly entraining, so pressure support could be
dominated by a thermal component not included in $P_{\rm min}$. 
In contrast, the lobes of FR\,IIs and the similar ``relaxed doubles''
have sharp boundaries \citeaffixed{LP91}{e.g.}, suggesting little
entrainment, as expected for cavities produced by powerful, relativistic
jets. This is confirmed by the observed holes in the ICM of 
Cygnus~A \cite{CPH94}. Only a handful of powerful sources have environments 
detectable with pre-{\em XMM\/}
X-ray observatories. We have observed two of them with {\em ROSAT\/}, and
discuss the results for these and others in the literature.
We assume $H_0 = 65 $ km\,s$^{-1}$ Mpc$^{-1}$ and $q_0=0$.


\section{Observations and results}

\begin{figure}
\begin{center}
\setlength{\unitlength}{1cm}
\begin{picture}(14,5.7)
\put(-0.1,-1.5){\includegraphics*[width=7.5cm]{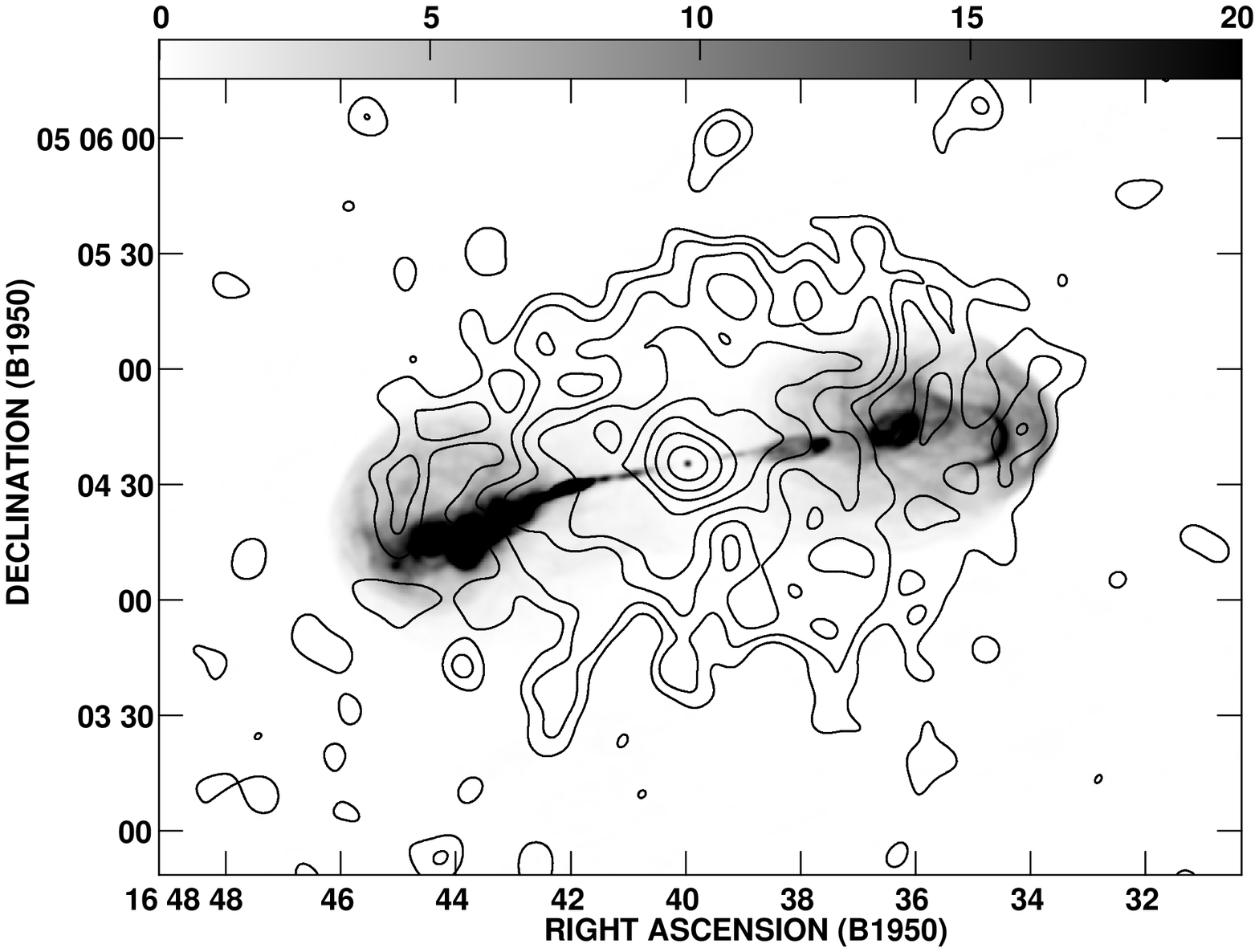}}
\put(7.7,5.5){\includegraphics*[angle=-90,width=6.5cm]{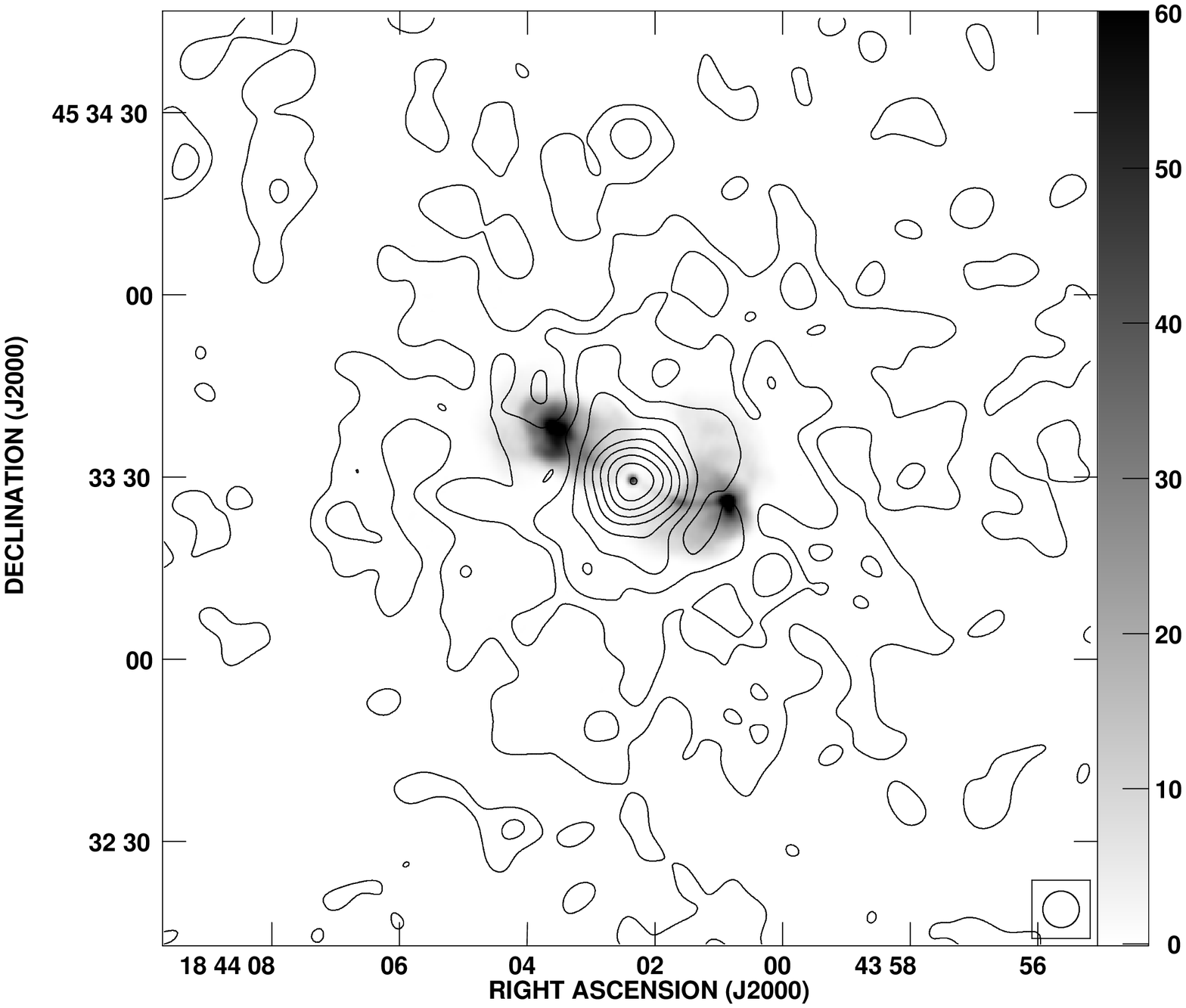}}
\end{picture}
\end{center}
\caption{Left: {\em ROSAT\/} HRI contours (separated by factors of 
$\sqrt{2}$) of the X-ray emission from Hercules A,
smoothed with a 10-arcsec gaussian, superimposed on a VLA $\lambda$20~cm
image from  Gizani \& Leahy (1999). 
Right: HRI contours (separated by intervals of 2$\sigma$) 
of 3C\,388, smoothed with a 6-arcsec gaussian, on
the VLA $\lambda$22~cm image from Roettiger et al. (1994).}
\label{overlays}
\end{figure}

We observed Hercules A  with the PSPC and HRI, 
and 3C\,388 with the HRI \cite{GL99,LG99}. 
Both are identified with cD galaxies at the 
centres of clusters. In both, the X-ray images (Fig.~\ref{overlays})
show compact central peaks, probably due to the AGN. 
We therefore fitted the X-ray radial profiles with a {\em ROSAT\/} 
point-spread function plus a modified King ($\beta$) model, giving
a reasonable fit in both cases, with good agreement between HRI and
PSPC for Her A.
 
We have also taken from the literature similar data for other cases
of powerful DRAGNs in cluster-centre galaxies.
All objects are more luminous than the
FR division and consist of sharply-bounded lobes, although only 3C\,295 and
Cyg~A are truly ``classical'' doubles.
Table~\ref{fits} lists for our full sample
the radio power at 178~MHz ($P_{178}$), the linear size ($D$),
and the parameters of the model fits to the X-ray data. 
We list ICM temperatures from {\em ROSAT\/} or {\em ASCA\/} spectra, 
except for 3C\,310 and 3C\,388 where we used the $L_X-T$ relation for 
clusters of galaxies.
We also list $u_{\rm min}/3$ for each lobe, found using the standard
formulae \citeaffixed{Leah91}{c.f.}. We assume no ``invisible'' energy
(e.g. in protons), a filling factor of unity, a 10~MHz spectral cutoff,
and an inclination of $50^{\circ}$ for the lobe axes. 
We used the spectral index $\alpha$ of the integrated emission below
1~GHz, where the lobes dominate.

\begin{table*}
\caption{Results of Minimum-energy calculations and $\beta$ model fits}
\label{fits}
\begin{center}
\begin{tabular}{lcccrrlcll}
\hline
DRAGN & $\lg(P_{178})$ & $\alpha$ & $u_{\rm min}/3$ & $D$ & $R_{\rm core}$ 
& $\beta$ & $n_0$ & $kT$ & Ref \\ 
 & W\,Hz$^{-1}$\,sr$^{-1}$ & & pPa & kpc  & kpc  & & $10^{3}$\,m$^{-3}$ 
&  keV &  \\
\hline
Hercules A & 27.33 & 1.01 & 0.30, 0.38 & 540 & 121 & 0.74 & ~6.5  & 2.45 & 1\\
3C\,388    & 25.68 & 0.70 & 0.72, 0.36 &  92 &  33 & 0.53 & 11   & 3  & 2,3 \\
Cygnus A   & 27.81 & 0.74 & 6.7, 8.6   & 147 &  41 & 0.75 & 65   & 4    & 4 \\
3C\,295    & 27.71 & 0.63 & 111, 112   &  34 &  42 & 0.56 & 40   & 7.1 & 5,6 \\
3C\,28     & 26.22 & 1.06 & 0.81, 0.84 & 153 &  50 & 0.67 & 20 & 4.9 & 7,8,9 \\
3C\,310    & 25.57 & 0.92 & 0.019, 0.028 & 340 & 84 & 0.5~ & ~2 & 2.5 & 7,10 \\
\hline
\end{tabular}
\end{center}
\medskip

\noindent
References: (1) Gizani \& Leahy (1999); (2) Leahy \& Gizani (1999);
(3) Roettiger et al. (1994); (4) Carilli et al. (1994); 
(5) Neumann (1999); (6) Leahy \& Spencer (1999);
(7) Hardcastle \& Worrall (1999); (8) Shibata et al. (1999).
(8) Leahy, Bridle \& Strom (1996); (10) Leahy \& Williams (1984). 
\end{table*}

%

\begin{figure}
\begin{center}
\includegraphics*[angle=-90,width=8cm]{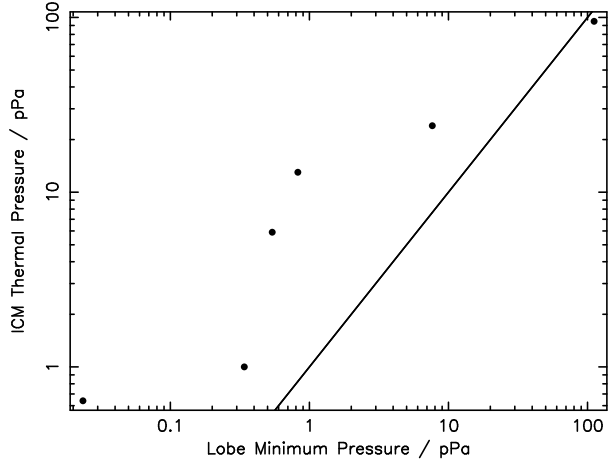}
\end{center}
\caption{Plot of the average (across the two lobes) minimum pressure 
against the ambient thermal pressure at a radius equal to two-thirds of
the distance from the core to the ends of the lobes.
The line shows $P_{\rm th} = P_{\rm min}$.} 
\label{meplot}
\end{figure}

Fig.~\ref{meplot} plots $P_{\rm min}$ against $P_{\rm th}$
at a ``typical'' (see caption) position for the lobe.
Clearly the two are related, but in most objects
$P_{\rm min} \simeq 0.1 P_{\rm th}$. The two objects with the highest
pressures and smallest discrepancy 
are the prototype classical doubles Cyg A and 3C\,295, which
are expected to be strongly {\em over-pressured},
which means that {\bf $P_{\rm min}$ is in general 
least an order of magnitude below the true lobe pressure}. 
Our sample is clearly
biased to high external pressures; but there is no reason why the ratio of
true to minimum pressure should depend on the actual value of the pressure, 
so the bias should not affect our conclusion.
The hotspot $P_{\rm min}$, not plotted here, is substantially 
above $P_{\rm th}$ for Cyg A and 3C\,295 but in 3C\,388 is still less than
$P_{\rm th}$. However in FR\,II models the whole lobe and not just the
hotspot is expected to be over-pressured.

\section{Discussion}

Inverse-Compton radiation (ICR) has now been detected from several radio lobes 
\cite{HCP94,FLKF95,Kana95,TLSB96,Tash98,BCSF99}.
The implied magnetic fields, $B$, are generally within 60\% of conventional 
minimum-energy values; 
there are no cases where detectable ICR is predicted for 
$B \lesssim B_{\rm me}$ but not seen. 
{\bf This suggests $B \approx B_{\rm me}$ in typical radio lobes}.
In contrast, our results would require
$B \approx 3 B_{\rm me}$ to retain equipartition, while numerical simulations
suggest that $B$ is below equipartition \citeaffixed{Clar93}{e.g.}. These
results could all be reconciled if the energy is dominated by
``invisible'' particles such as protons or low-energy electrons.
For 3C\,388, with its
rather flat low-frequency spectra, it is not even enough to extend
the energy spectrum to Lorentz factor $\gamma =1$; however $e^-$-$e^+$
jets may be saved if the observed power-law is only a ``tail''
on a relativistic-Maxwellian energy distribution (the lobe temperature must
be relativistic if fed by relativistic jets).

\citeasnoun{RS91} used minimum-energy estimates and spectral
ages to derive a
relation between jet power, $Q$, and narrow-line luminosity in 3CR 
radio galaxies: $Q \simeq 100 L_{\rm NLR} \approx 550 
L_{\mbox{\scriptsize [O{\sc iii}]}}$.
Similarly, \citeasnoun{FMB93} relate
accretion disc luminosity (effectively the big blue bump) to [O{\sc iii}]
``magnitude'' for PG quasars, which can be decoded as:
$L_{\rm disc} \simeq 430 L_{\mbox{\scriptsize [O{\sc iii}]}}$.
The Falcke et al. relation appears to hold irrespective of radio-loudness.
\citeasnoun{JB90} show that [O{\sc iii}] is somewhat anisotropic,
with quasars 5--10 times more luminous in [O{\sc iii}] than 
corresponding radio galaxies;
however, as the Rawlings \& Saunders correlation is partly based on the
more isotropic [O{\sc ii}], we estimate a factor of 3 offset between
the effective [O{\sc iii}] luminosities in the above two relations. 
To be conservative,
we assume that the AGN bolometric luminosity $L_{AGN}$ is twice that
of the big blue bump, to allow for the IR and X-ray peaks. Thus for
radio loud objects, the Rawlings \& Saunders result implies
$Q \simeq 0.2 L_{AGN}$.
A primary uncertainty in the spectral age, the use of $B_{\rm me}$,
is strongly supported by the ICR results. A systematic underestimate of
ages of up to about 3 is possible from mixing within the lobes \cite{Trib93}. 
Even allowing for that, 
{\bf our calibration of $u_{\rm min}$ implies that
the jet kinetic luminosity in radio-loud AGN
is at least as large as the bolometric luminosity of the AGN, and
may substantially exceed it}. 
There is no reason why this should not be so, 
as jets and thermal radiation are independent by-products of accretion;
jets are not powered by the AGN radiation.

If jets are a primary energy loss mechanism of AGN in cluster-centre
ellipticals (which are always radio-loud if active at all), 
then much of the accretion energy produced by the growth of the 
central black hole does work on the cluster gas, heating it.
Taking $L \approx 0.1\dot{M}c^2$, we assume half of this goes into
the jet, and (consistent with Rawlings and Saunders) that half of the
jet power does work on the cluster as opposed to being ultimately radiated
as very-low-frequency radio waves, the total energy supplied to the cluster
gas is $0.025 M_{\bullet}c^2$. If central black holes really follow the
\citeasnoun{Mago98} relation, this corresponds to $1.5\times 10^{-4}$
times the galaxy rest mass, more than ten times
the available energy from supernova-driven winds
\citeaffixed{PCN99}{c.f.}. Thus jets must have had a dramatic effect 
on the cluster gas at early
epochs, where most AGN activity and, presumably, black hole growth 
occurred.

\end{document}